\documentstyle[12pt,epsfig]{article} 

\def\NPB#1#2#3{Nucl.\ Phys.\ {\bf B#1}, #3 (19#2)}

\def\PLB#1#2#3{Phys.\ Lett.\ {\bf B#1}, #3 (19#2)}

\def\PRD#1#2#3{Phys.\ Rev.\ {\bf D#1}, #3 (19#2)}
\def\PRL#1#2#3{Phys.\ Rev.\ Lett.\ {\bf#1}, #3 (19#2)}

\def\MPLA#1#2#3{Mod.\ Phys.\ Lett.\ {\bf A#1}, #3 (19#2)}

\def\EPJC#1#2#3{Eur.\ Phys.\ J.\ {\bf C#1}, #3 (19#2)}
\def\PTP#1#2#3{Prog.\ Theor.\ Phys.\ {\bf#1}, #3 (19#2)}
\newcommand{\newc}{\newcommand}
\newcommand{\gsim}{ \mathop{}_{\textstyle \sim}^{\textstyle >} }
\newcommand{\lsim}{ \mathop{}_{\textstyle \sim}^{\textstyle <} }

\newc{\gev}{\,{\rm GeV}}
\newc{\hl}{{h^0}}
\newc{\hh}{{H^0}}
\newc{\hA}{{A}}
\newc{\rbtau}{R_{b/\tau}}
\newc{\rbc}{R_{b/c}}
\newc{\rbtaua}{R^{A}_{b/\tau}}
\newc{\mz}{m_Z}
\newc{\mw}{m_W}
\newc{\mx}{m_{GUT}}
\newc{\beq}{\begin{equation}}
\newc{\eeq}{\end{equation}}
\newc{\bea}{\begin{eqnarray}}
\newc{\eea}{\end{eqnarray}}
\newc{\ie}{{\it i.e.\/}}
\def\vev#1{\left\langle #1 \right\rangle}

\begin{document}
%%%%%%%%%%%%%%%%%%%%%%%%%%%%%%%%%%%%%%%%%%%%%%%%%%%%%%%%%%%%%%
\catcode`@=11
% Redefine caption to put text and formulas in smaller font
\long\def\@caption#1[#2]#3{\par\addcontentsline{\csname
  ext@#1\endcsname}{#1}{\protect\numberline{\csname
  the#1\endcsname}{\ignorespaces #2}}\begingroup
    \small
    \@parboxrestore
    \@makecaption{\csname fnum@#1\endcsname}{\ignorespaces #3}\par
  \endgroup}
\catcode`@=12
%%%%%%%%%%%%%%%%%%%%%%%%%%%%%%%%%%%%%%%%%%%%%%%%%%%%%%%%%%%%

\begin{titlepage}
\def\thefootnote{\fnsymbol{footnote}}
\begin{flushright}
{\rm
OSU--HEP--98--09\\
LBNL--42521\\
hep-ph/9811308\\
November 1998
}
\end{flushright}
\vskip 1cm

\begin{center}
{\Large \bf
Signatures of Supersymmetry and
Yukawa \\[2mm]
Unification in Higgs Decays
}
\vskip 0.7cm
{\large
K.S.~Babu$\,{}^1$\footnote{Email: babu@osuunx.ucc.okstate.edu} 
{\normalsize and} Christopher Kolda$\,{}^2$\footnote{Email: CFKolda@lbl.gov}\\}
\vskip 0.4cm
{\small\em ${}^1$ Department of Physics, Oklahoma State University,
  Stillwater, OK 74078, USA
 \\[2mm]
${}^2$ Theoretical Physics Group, MS~50A-5101, Lawrence Berkeley National
Laboratory\\
University of California, Berkeley, CA 94720, USA\\
}
\end{center}
\vskip .5cm

\begin{abstract}

We show that the branching ratio $\rbtau \equiv
BR(h^0\to\bar bb)/BR(h^0\to\tau^+\tau^-$) of the Higgs boson $h^0$
may usefully differentiate
between the Higgs sectors of the Minimal Supersymmetric Standard Model (MSSM)
and non--supersymmetric models such as the Standard Model or its 
two Higgs doublet extensions.  Although at tree level $\rbtau$ is the same
in all these models, only in the MSSM can it receive a large radiative
correction,
for moderate to large values of the parameter $\tan\beta$.  Such large
corrections are motivated in supersymmetric unified schemes wherein the Yukawa
couplings of the $b$--quark and the $\tau$--lepton are equal at the unification
scale; otherwise the $b$--quark mass prediction is too large by 15--30\% for
most of parameter space.  Thus accurate measurements of the Higgs branching
ratios can probe physics at the unification scale. The branching ratio of
$h^0$ into charm quarks, as well as of the other Higgs bosons 
$(H^0,A^0)$
into $\bar bb$, $\tau^+\tau^-$, $\bar cc$ can provide additional information
about the supersymmetric nature of the Higgs sector.

\end{abstract}

\end{titlepage}
\setcounter{footnote}{0}
\setcounter{page}{1}

\section{Introduction}

The symmetry breaking sector of the Standard Model (SM) is
still being vigorously pursued.
The Minimal Supersymmetric Standard Model (MSSM) is perhaps the most
widely anticipated, for it can explain naturally why the Higgs boson remains
light;
it is also compatible with the unification of the three gauge couplings.

The Higgs sector of the MSSM is a
special case of the more general two--Higgs doublet Standard Model (2HDSM) with
three characteristic features~\cite{hhg}.
First, as in the generic 2HDSM, the mass of the lightest CP--even Higgs is
controlled by dimensionless couplings; however, the dimensionless
couplings which appear in the MSSM Higgs potential are simply
the SU(2)$\times$U(1) gauge couplings.
Thus one derives the remarkable (tree--level) result $m_h<\mz$.
Second, the Higgs potential of the MSSM can be made real by appropriate field
redefinitions and gauge transformations
so that, again at tree level, there is no CP violation arising from the Higgs
sector itself.

Finally, the MSSM is a so--called ``type II'' model wherein one Higgs
doublet (denoted $H_u$) couples to the $u$--quarks, while the second Higgs
doublet ($H_d$) couples to the $d$--quarks and charged leptons.
Such a division of the fields is a requirement in
supersymmetric (SUSY) models imposed by holomorphy and anomaly cancellation.
But there are other
incidental benefits. For example, the large flavor--changing neutral currents
endemic to the general 2HDSM are
avoided. There is one other side--effect as well: the ratios of branching
ratios
for a Higgs boson
decaying into quarks and leptons in the same class should match the ratio
calculated in the SM. This last result implies, for example, that $\rbtau\equiv
BR(\hl\to\bar bb)/BR(\hl\to\tau^+\tau^-)$ is the same in MSSM as in
the SM, roughly $3m_b^2/m_\tau^2$.
Likewise for the ratio of $t$--quarks to $c$--quarks, or $b$--quarks to
$s$--quarks, and so on. The invariance of these double ratios
has long been known to be a distinguishing feature of any type II 2HDSM, and
the MSSM in particular.

  All three of the above results can receive significant alterations due to
radiative corrections. The mass of the lightest Higgs receives
substantial corrections from the heavy $t$--quark, increasing its upper
bound to roughly $130\gev$~\cite{hmass}.
CP violation can also enter the Higgs
couplings through spontaneous CP violation in the one--loop effective
potential (this turns out to be typically very small \cite{maekawa}) or finite
corrections to the Higgs--matter couplings~\cite{bkmw}.
In this paper we will show that such
finite corrections may also significantly shift the ratios of branching ratios,
such as $\rbtau$,
in interesting regions of SUSY parameter space. Such shifts will {\it not}
occur in the SM or in the non--SUSY 2HDSM and so can serve as an
indicator of SUSY.  The double ratio $\rbc$, of the Higgs branching ratios
into $\bar bb$ versus $\bar cc$, can provide additional
information, as can the branching ratios of the other two neutral 
Higgs bosons.  We will also explain how such shifts
may provide a new experimental handle on models of grand unification.

It is quite conceivable that a light Higgs boson will be discovered before
any supersymmetric particles.  If the branching ratio
$\rbtau$ of the Higgs is measured to be significantly different
from the SM prediction, our results suggest,
it would be a strong indication of the supersymmetric
nature of the Higgs sector.

\section{The MSSM Higgs sector at tree level}

The coupling of the Higgs multiplets to the usual SM fermions is
described in a SUSY model via the superpotential:
\beq
W= y_u Qu^cH_u+y_d Qd^cH_d+y_e Le^cH_d+\mu H_uH_d
\eeq
where the $y_i$ are the Yukawa couplings ($3\times3$ matrices in
generation space), and $\mu$ is a SUSY--invariant mass parameter which
mixes the two Higgs doublets.
After electroweak symmetry breaking, the fermions
get masses at tree--level of, for example:
\beq
m_t=y_t v \sin\beta, \quad \quad m_{b,\tau}=y_{b,\tau} v \cos\beta
\eeq
where $\tan\beta\equiv\vev{H_u}/\vev{H_d}$ and
$v^2=\vev{H_u}^2+\vev{H_d}^2\simeq(174\gev)^2$.
Perturbativity usually constrains $\tan\beta$ to lie in the range
$1.4\lsim\tan\beta\lsim60$.

Because $SU(2)\times U(1)$ is broken, the
interaction eigenstates $H_u$ and $H_d$ also mix. The spectrum of
the Higgs sector is then described by 3 Goldstone bosons eaten by
$W^\pm$ and $Z^0$, a pair of charged Higgs $H^\pm$, a neutral pseudoscalar
$A^0$, and two neutral scalars $\hl$ and $\hh$, the latter defined so that
$m_h<m_H$. The mass eigenstates for the two neutral scalars are
defined via:
\beq
\left(\begin{array}{c} \hl \\ \hh \end{array}\right)
=\sqrt{2}
\left(\begin{array}{cc} \cos\alpha & -\sin\alpha \\ \sin\alpha &
\cos\alpha \end{array}\right)
\left(\begin{array}{c} \mbox{Re}\,H_u^0 \\ \mbox{Re}\,H_d^0
  \end{array} \right) \ .
\eeq
The pseudoscalar Higgs is the combination $A^0 = \sqrt{2}(\cos\beta\,
\mbox{Im}\,H_u^0 +\sin\beta\,\mbox{Im}\,H_d^0)$.
Given the two mixing angles $\alpha$ and $\beta$, the couplings of the
quarks and leptons are completely determined in terms of their SM
values. One finds that, for the lighter scalar, the ratio of the MSSM
couplings to those of the SM are simply:
\beq
\frac{\hl\bar tt|_{\rm MSSM}}{\hl\bar tt|_{\rm SM}}=\frac{\cos\alpha}
{\sin\beta}\ , \quad\quad\quad
\frac{\hl\bar bb|_{\rm MSSM}}{\hl\bar bb|_{\rm SM}}=
\frac{\hl\tau^+\tau^-|_{\rm MSSM}}{\hl\tau^+\tau^-|_{\rm SM}}=
\frac{-\sin\alpha}{\cos\beta}\ ,
\eeq
where the SM couplings are $g_2 m_f/2\mw$ for fermion $f$.
Note that since the $b$--quark and $\tau$--lepton both couple to the
same Higgs interaction eigenstate, their couplings to the physical
Higgs bosons are both shifted by the same amount. Therefore the ratio
of branching ratios $\rbtau\equiv
BR(\hl\to\bar bb)/BR(\hl\to\tau^+\tau^-)$ is the same in the SM and
MSSM, namely $3m_b^2/m_\tau^2$ up to kinematic factors and Standard
Model QCD corrections (we ignore the small QED and electroweak
corrections)~\cite{gorishny}:
\beq
\rbtau= 3\,
\frac{m_b^2}{m_\tau^2}\,\left(\frac{m_h^2
-4m_b^2(m_b)}{m_h^2-4m_\tau^2(m_\tau)}\right)^{1/2}
\left[1 + 5.67\,\frac{\alpha_s(m_h)}{\pi}+29.14\,\frac{\alpha_s^2(m_h)}
{\pi^2}\right] \ .
\label{sm}
\eeq
(For this letter, $m_b$ and $m_\tau$ are to be evaluated in the $\overline{MS}$
scheme and at the scale
$Q=m_h$ unless otherwise specified.
In our numerical calculations we take $Q=\mz$;
we are then missing only a very small residual correction proportional to
$\log(m_h/\mz)$.) Defining $\omega$ by
$\rbtau=3(m_b^2/m_\tau^2) (1+\omega)$ one finds,
for $\alpha_s(\mz^2)=0.119$,
that $\rbtau$ receives a QCD/phase space enhancement over its tree--level
value of $(1+\omega)\simeq1.25$.

\section{Finite corrections at one--loop}

\begin{figure}
\centering
\epsfysize=1.5in
\epsffile{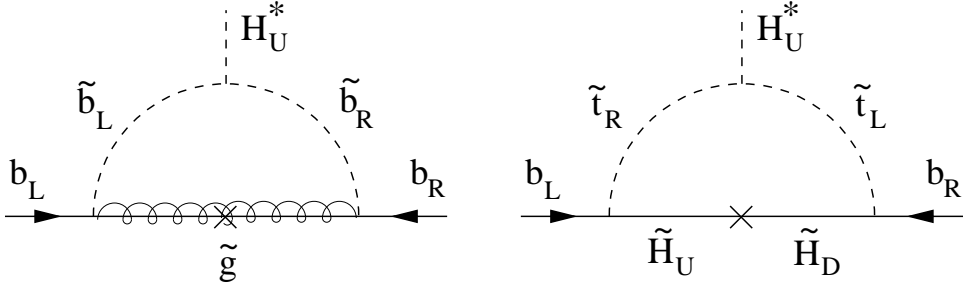}
\caption{Leading threshold contributions to $\epsilon_b$.}
\label{fig:fdiag}
\end{figure}

Although the $b$--quark does not couple to $H_u^0$ at tree level, it picks up
a small coupling to
$H_u^0$ at one--loop through the diagrams in Fig.~\ref{fig:fdiag}.
Such diagrams were studied earlier in the context of neutron electric dipole
moments
in Ref.~\cite{edm} and in the context of radiative fermion mass generation
in Ref.~\cite{banks}.
Their existence is due to the interesting fact that $b$--squarks {\em do}
couple to $H_u^0$, despite the fact that $b$--quarks do not, through the
interaction $y_b\mu\widetilde b_L\widetilde b_R^* H_u^{0*}+h.c.$\
Since the coupling is loop--suppressed it is small. However, if $v_u\gg v_d$
(\ie, $\tan\beta$ is large), the contribution of these loops to the $b$--mass
is
enhanced by $\tan\beta$ and can therefore become quite significant.
%%%%%%%%%%%%%%%%%%%%%%%%%%%%%%%%%%%%%%%%%%%%%%%%%%%%%%%%%%%%%%%%%%%
%%%%%%%%%%%%%%%%%%%%%%%%%%%%%%%%%%%%%%%%%%%%%%%%%%%%%%%%%%%%%%%%%%%

The significance of such finite
corrections for Yukawa unification schemes in grand unified
theories was noticed in Ref.~\cite{hrs}. Specifically, in a class of
minimal SO(10) models, one expects unification of all third generation
Yukawas at the GUT scale: $y_t(\mx)=y_b(\mx)=y_\tau(\mx)$. Such a
unification demands large $\tan\beta$ to compensate the large
hierarchy in the masses of the $b$-- and $t$--quarks: $m_t/m_b=(y_t/y_b)
\tan\beta$. In the context of these unified models, it was realized
that the one--loop contribution from Fig.~\ref{fig:fdiag}
could significantly shift the
$b$--mass, obscuring the simple relation between $m_b$ and $y_b$. This
should become clear shortly.

Begin by writing the most general coupling of a $b$--quark and $\tau$--lepton
to
the neutral Higgs fields:
\beq
{\cal L}=y_b \bar b_L b_R H_d^0 + \epsilon_b y_b \bar b_L b_R H_u^{0*} +
(b\leftrightarrow\tau) + h.c.
\eeq
where we assume for simplicity that $y_{b,\tau}$ and $\epsilon_{b,\tau}$ are
all real. For the remainder of this letter we will also assume that
the scale of SUSY--breaking is larger than that of
electroweak--breaking, consistent with our experimental knowledge of SUSY.
$\epsilon_b$ receives contributions, after electroweak-- and
SUSY--breaking, from a number of diagrams, the most important being
those in Fig.~\ref{fig:fdiag}, which yield (taking all parameters real):
\beq
\epsilon_b=\frac{2\alpha_3}{3\pi} \mu M_3 f(M_3^2,m^2_{\widetilde b_L},
m^2_{\widetilde b_R})+\frac{y_t^2}{16\pi^2} \mu A_t f(\mu^2,
m^2_{\widetilde t_L},m^2_{\widetilde t_R})
\eeq
where
\beq
f(m_1^2,m_2^2,m_3^2)\equiv \frac{1}{m_3^2}\left[\frac{x\log x}{1-x}-\frac{y\log
y}{1-y}\right]\frac{1}{x-y}\
\eeq
for $x=m_1^2/m_3^2$, $y=m_2^2/m_3^2$.  Numerically $\epsilon_b$ is of order
2\%, though the precise value will depend on the supersymmetric spectrum.
There are also contributions with internal $SU(2)\times U(1)$ gauginos, but
these are typically suppressed by $\alpha_{1,2}/\alpha_3$ compared to
the gluino diagram\footnote{If there is a significant
contribution to $\epsilon_b$ coming from diagrams other those of
Fig.~\protect\ref{fig:fdiag}, then these can be simply absorbed into
$\epsilon_b$ and the rest of our discussion is unchanged.}.
Notice that ${\rm sgn}(\epsilon_b)$ is model--dependent and cannot be
predicted without further input. Since there are no QCD--enhanced
contributions for the $\tau$, nor a light right--handed $\nu_\tau$, then
to a good approximation $\epsilon_\tau\simeq0$;
we will comment later on the possibility of $\epsilon_\tau\neq0$.

Including the corrections, the $b$--quark mass can be written
\beq
m_b=y_b v_d + y_b\epsilon_b v_u = y_b (1 + \epsilon_b \tan\beta) v\cos\beta\ .
\eeq
Meanwhile, the coupling of the $b$--quark to the light Higgs is given by:
\beq
{\cal L}_{h\bar bb}=\frac{1}{\sqrt{2}}
(-y_b\sin\alpha+y_b\epsilon_b\cos\alpha)h^0 \bar bb\ .
\eeq
Including the corrections, the ratio of branching ratios, $\rbtau$, can be
expressed (including the usual phase space/QCD corrections) as:
\begin{eqnarray}
\rbtau &=& 3\,\frac{y_b^2}{y_\tau^2}\,(1-\epsilon_b/\tan\alpha)^2
(1+\omega) \nonumber \\
&=& 3\,\frac{m_b^2}{m_\tau^2}
\frac{(1-\epsilon_b/\tan\alpha)^2}{(1+\epsilon_b\tan\beta)^2}(1+\omega)\ .
\label{rbtau-full}
\end{eqnarray}

A couple of comments are now in order. First, perturbativity requires that
$\epsilon_b\ll1$, though not necessarily $\epsilon_b\tan\beta\ll1$. Thus, the
$b$--quark mass can receive a significant correction from
Fig.~\ref{fig:fdiag}. Second, the
values of $\alpha$ and $\beta$ are correlated in the MSSM via:
\beq
\sin2\alpha=-\frac{m^2_\hA+\mz^2}{m^2_H-m^2_h}\sin2\beta
\simeq-\frac{m^2_\hA+\mz^2}{|m^2_\hA-\mz^2|}\,\sin2\beta\
\eeq
where the final approximation
holds in the large $\tan\beta$ limit. (We use the exact relation
between $\alpha$ and $\beta$ in our numerical results.)
Thus in the so--called ``Higgs decoupling limit'' of the MSSM, in which
$m_\hA\to\infty$, one finds $\alpha\to\beta-\pi/2$ and so $\rbtau$
approaches its SM value, given by Eq.~(\ref{sm}). Thus we expect to see the
largest deviations of
$\rbtau$ from its SM value for relatively light $A^0$. We can expand $\rbtau$
in the limit of $m_\hA\gg\mz$:
\beq
\rbtau\simeq 3\frac{m_b^2}{m_\tau^2}(1+\omega)\left\{1-4\frac{\mz^2}{m_\hA^2}\,
\frac{\epsilon_b\tan\beta}{1+\epsilon_b\tan\beta}\right\}\ .
\label{maform}
\eeq
In this form, the shift in $\rbtau$ away from its SM value,
$\delta\rbtau$, can be written as a function only of the shift in the
$b$--mass coming at one--loop, $\delta m_b/m_b=\epsilon_b\tan\beta$:
\beq
\frac{\delta\rbtau}{\rbtau}\simeq-4\frac{\mz^2}{m_\hA^2}\,
\frac{\delta m_b/m_b}{1+\delta m_b/m_b}\ .
\label{dmform}
\eeq
(If there is a significant contribution to $\epsilon_\tau$, then so
long as $\epsilon_\tau\tan\beta\ll1$, one can simply replace
$\epsilon_b$ with  $(\epsilon_b-\epsilon_\tau)$ in Eq.~(\ref{maform}) and
corresponding formulae, and replace $\delta m_b/m_b$ with $(\delta
m_b/m_b - \delta m_\tau/m_\tau)$ in Eq.~(\ref{dmform}) and corresponding
formulae. If $\epsilon_\tau\tan\beta\sim1$, similarly simple forms occur.)

\begin{figure}[t!]
\centering
\epsfysize=5in
\epsffile{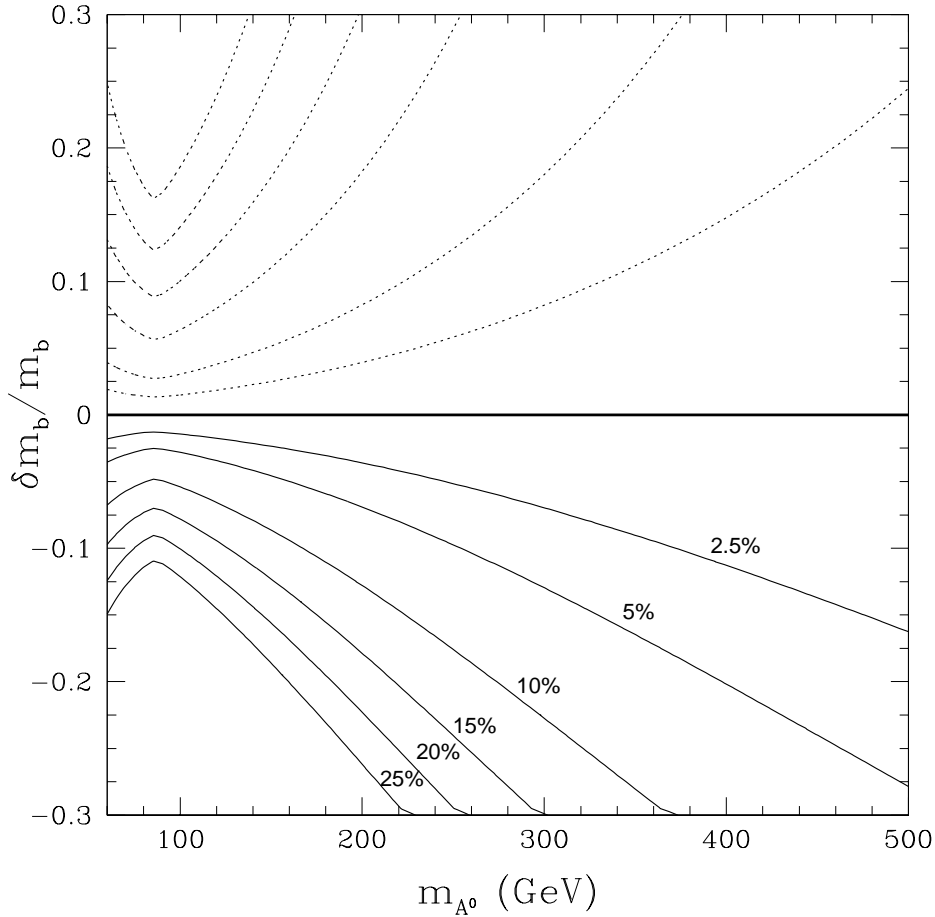}
\caption{Contours of $\delta\rbtau/\rbtau$ in the $m_\hA - \delta
  m_b/m_b$ plane. The central dark contour is $\delta\rbtau/\rbtau=0$;
dotted contours are negative values of the corresponding labelled
solid contours.
The figure does not include the leading radiative corrections to the
Higgs mass matrix, which are discussed in the text. The figure holds
for all $\tan\beta\gsim5$.
}
\label{fig:hdecays}
\end{figure}

In Fig.~\ref{fig:hdecays}
we have shown contours of $\delta\rbtau/\rbtau$ in the $m_\hA$ --
$\delta m_b/m_b$
plane; the plot uses the full result of Eq.~(\ref{rbtau-full}) for
$\tan\beta=30$, though the dependence of the plot on $\tan\beta$ is
unobservable for any
$\tan\beta\gsim5$. Contours are shown for $\delta\rbtau/\rbtau$ of +2.5, 5, 10,
15, 20, 25\% (solid lines in lower half plane), and -2.5, -5, -10, -15, -20,
-25\%
(dotted lines in upper half
plane). Even though the plot has no apparent dependence on $\tan\beta$ it is
unlikely that any large effect would be observed at small values of
$\tan\beta$. That is to say, although the $\tan\beta$ dependence can
be absorbed into $\delta
m_b/m_b$, it would be very difficult to generate appreciable shifts in the
$b$--quark mass without the enhancement provided by large $\tan\beta$.
Typically,
one would require $\tan\beta \gsim 10$ or so, for $ \delta m_b/m_b \sim
20\%$.
%%%%%%%%%%%%%%%%%%%%%%%%%%%%%%%%%%%%%%%%%%%%%%%%%%%%%%%%%%%%%%%%%%%
%%%%%%%%%%%%%%%%%%%%%%%%%%%%%%%%%%%%%%%%%%%%%%%%%%%%%%%%%%%%%%%%%%%

It is well--known that radiative corrections in the MSSM coming from
heavy top squark loops can significantly alter the scalar Higgs mass
matrix, lifting the lighter scalar above the $Z$ mass. These same
corrections alter the relation between $\alpha$ and $\beta$, slowing
the decoupling that occurs as $m_\hA\to\infty$. Each of the elements
of the scalar Higgs mass matrix is shifted by corrections, but for the
sake of simplicity, we will only keep the leading term which shifts
the diagonal $H_u^0$ piece by an amount~\cite{hmass}:
\beq
\delta m^2=\frac{3 g^2 m_t^4}{8\pi^2\mw^2}\log\left(\frac
{m^2_{\widetilde t}}{m_t^2}\right)\ .
\eeq
At large $\tan\beta$, one measures $\delta m^2$ simply by
discovering the light Higgs, since $\delta m^2\simeq m^2_h-\mz^2$ to
a good approximation.
The shifted $\alpha$ is now given by:
\beq
\sin2\alpha=-\frac{(m^2_\hA+\mz^2)}{\Delta}\sin2\beta
\label{radcorr}
\eeq
where
\bea
\Delta&=&
\left[(m^2_\hA+\mz^2+\delta
  m^2)^2 - 4m^2_\hA\mz^2\cos^2 2\beta \right.\nonumber \\
 & &\left. {}-4m^2_\hA\delta m^2\sin^2\beta
  -4\mz^2\delta m^2\cos^2\beta\right]^{1/2}\ .
\eea
In the large $\tan\beta$ limit,
Eq.~(\ref{radcorr}) simplifies to
\beq
\sin2\alpha\simeq-\frac{m_A^2+\mz^2}{|m_A^2-m_h^2|}\sin2\beta
\simeq-\left(1+\frac{2\mz^2+\delta m^2}{m^2_\hA}\right)\sin2\beta
\eeq
where the last equality holds if we also take the large $m_A$ limit.
These show clearly that for $\delta m^2>0$ the radiative corrections slow
down the decoupling as $m_\hA\to\infty$. Thus Eq.~(\ref{dmform}) is
corrected by replacing $\mz^2$ with $(\mz^2+\delta m^2/2)$. For
$m_h=120\gev$, for example, the radiative corrections increase
$|\delta\rbtau/\rbtau|$ from the values shown in
Fig.~\ref{fig:hdecays} by roughly 40\%.  (That is, the curve labeled
10\% would correspond to 14\% after radiative correction.)  There are
additional, but smaller,
corrections to the other entries in the scalar Higgs mass matrix. Though these
corrections have the ability to shift the mixing angle, $\alpha$, they are more
model--dependent and we do not analyze them here; see however Refs.~\cite{lw}
for
a discussion of some possible effects.

We can define another double ratio $\rbc \equiv BR(h^0\to\bar bb)
/BR(h^0\to\bar cc)$.  In the MSSM, including the finite
radiative corrections, this is given by
\begin{equation}
\rbc = { m_b^2 \over m_c^2 }\, (\tan\alpha\tan\beta)^2 \left[
{1-\epsilon_b/\tan\alpha \over 1 + \epsilon_b\tan\beta}\right]^2
\left[{1 + \epsilon_c/\tan\beta \over 1 -
    \epsilon_c\tan\alpha}\right]^2\ ,
\label{rbc}
\end{equation}
where $\epsilon_c y_c$ is the radiatively generated coefficient to the
$H_d^{0*}\bar cc$ vertex.
Note that there is no $\tan\beta$ enhancement associated with $\epsilon_c$
($\epsilon_c \simeq \epsilon_b \sim 2\%$ from the gluino graph), so
the rightmost bracket in Eq.~(\ref{rbc}) goes to 1 and 
$\delta\rbc/\rbc$ becomes identical to
$\delta\rbtau/\rbtau$.  Thus simultaneous measurement of
a shift in $\rbc$ will
provide crucial supporting evidence for supersymmetry.

\begin{figure}
\centering
\epsfysize=2.75in
\epsffile{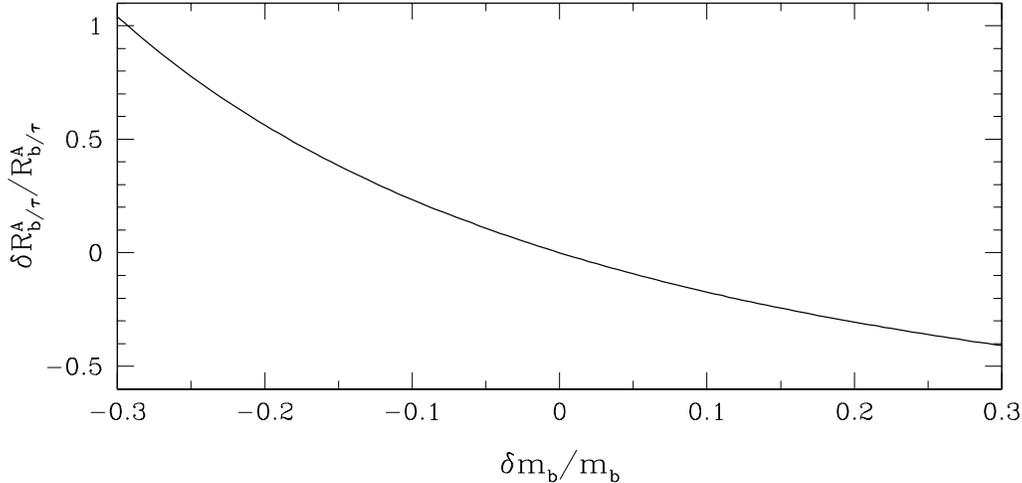}
\caption{Corrections to $\rbtaua$ as a function of $\delta
  m_b/m_b$. The figure holds for all $\tan\beta\gsim5$.}
\label{fig:adecays}
\end{figure}

The same analysis can be repeated for the heavier scalar Higgs, $\hh$, simply
by replacing $-1/\tan\alpha\to\tan\alpha$ in Eq.~(\ref{rbtau-full}),
and for the pseudoscalar, $A^0$, by replacing $\tan\alpha\to-\tan\beta$ in the
numerator of Eq.~(\ref{rbtau-full}). Note that
for the $\hh$ and $A^0$ decoupling does not occur, as expected.
At large $\tan\beta$, the expressions for $\hh$ and $A^0$
simplify to the same form,
so that the shift in either ($\equiv\rbtaua$) can be written:
\beq
\delta\rbtaua/\rbtaua\simeq \left(1+\delta m_b/m_b\right)^{-2}-1\ .
\eeq
Because there is no dependence on $m_\hA$ (nor in this form on $\tan\beta$),
the result is particularly easy to examine. In Fig.~\ref{fig:adecays}
we do just that, showing
the shift in $\rbtaua$ as a function of the shift in the $b$--quark
mass. Notice that a 25\% shift in the $b$--mass translates into a 75\%
correction to $\rbtaua$!
%%%%%%%%%%%%%%%%%%%%%%%%%%%%%%%%%%%%%%%%%%%%%%%%%%%%%%%%%%%%%%%%%%%
%%%%%%%%%%%%%%%%%%%%%%%%%%%%%%%%%%%%%%%%%%%%%%%%%%%%%%%%%%%%%%%%%%%

\section{Implications for SUSY searches}

It is entirely conceivable, if not likely, that the lightest Higgs will be
discovered prior to the discovery of any SUSY partners. It is then a fair
question to ask: is this Higgs the SM Higgs, a SUSY Higgs, or some other? If
the Higgs is found to have appreciable deviations in $\rbtau$ from the SM case,
though not 100\%, we believe that this will be a fair argument that SUSY exists
and will be found at higher energies. In the SM itself, there is no source of
large corrections to $\rbtau$ besides those already shown in
Eq.~(\ref{sm}). In non--SUSY
extensions of the SM Higgs sector,
there is no simple source for {\it large} corrections to
$\rbtau$. For example, in the type II 2HDSM, there is a diagram with a top
quark and charged Higgs
boson which corresponds closely to the second diagram of Fig.~\ref{fig:fdiag},
it can be
shown to lack the $\tan\beta$ enhancement that the SUSY diagrams
share~\cite{hrs}, making it very difficult to generate large enough $\delta
m_b/m_b$ to be observable.

We should remark that there are ways to
distinguish the MSSM Higgs sector from other (perhaps less motivated)
versions of the non--SUSY
2HDSM by studying the decays of $h^0$ alone.
For example, in the type I model, where all the fermions couple
to a single Higgs, the predictions for $\rbtau$ will be identical
to that of the SM.  In the type III model, where $H_u$ couples to $u$--quarks
and charged leptons while $H_d$ couples to $d$--quarks, already at
the tree level $\rbtau$ is different from its SM value.  This case
can be tested by the measurements of three observables: $\rbtau$, $\rbc$ and
$\sigma \cdot BR(\hl\to\bar bb)$, where $\sigma$ denotes the
production cross section for $\hl$.  Since there are no large radiative
corrections in this model, these three observables depend only on two
parameters, {\it viz.}, $\alpha$ and $\beta$ (apart from $m_h$),
so consistency of this scenario can be directly tested.

Even if there are early indications of SUSY in the Higgs
decays, this would not be equivalent to
saying that SUSY is light. On the contrary, the diagrams that contribute to
$\epsilon_b$ are non--decoupling --- as the SUSY scale increases, these
diagrams
approach a non--zero constant, so long as $m_A$ remains light.
Thus corrections to Higgs decays widths may be
the only indication of SUSY for quite some time. This could be a
particularly interesting probe then of
SUSY models in which much of the SUSY spectrum remains quite heavy.

\section{Implications for grand unification}

The simplest and most elegant grand unification theories (GUTs), SU(5) and
SO(10),
group otherwise different fermions into common representations at the
unification scale. For example, in SU(5) the $b_R$ and $\tau_L$ are part of a
single ${\bf\bar5}$ representation, while the $b_L$ and $\tau_R$ are part of a
single ${\bf10}$; in SO(10), all the above are grouped together into a single
${\bf16}$ spinor representation. For minimal GUT Higgs sectors, the grouping
imply $y_b=y_\tau$ for SU(5) and $y_t=y_b=y_\tau$ for SO(10), all evaluated at
the unification scale~\cite{ceg}. In most of the simplest extensions of the GUT
Higgs
sectors, the $b-\tau$ unification of SU(5) and SO(10) survives, though not
always the full $t-b-\tau$ unification of SO(10).

\begin{figure}[t!]
\centering
\epsfysize=5in
\epsffile{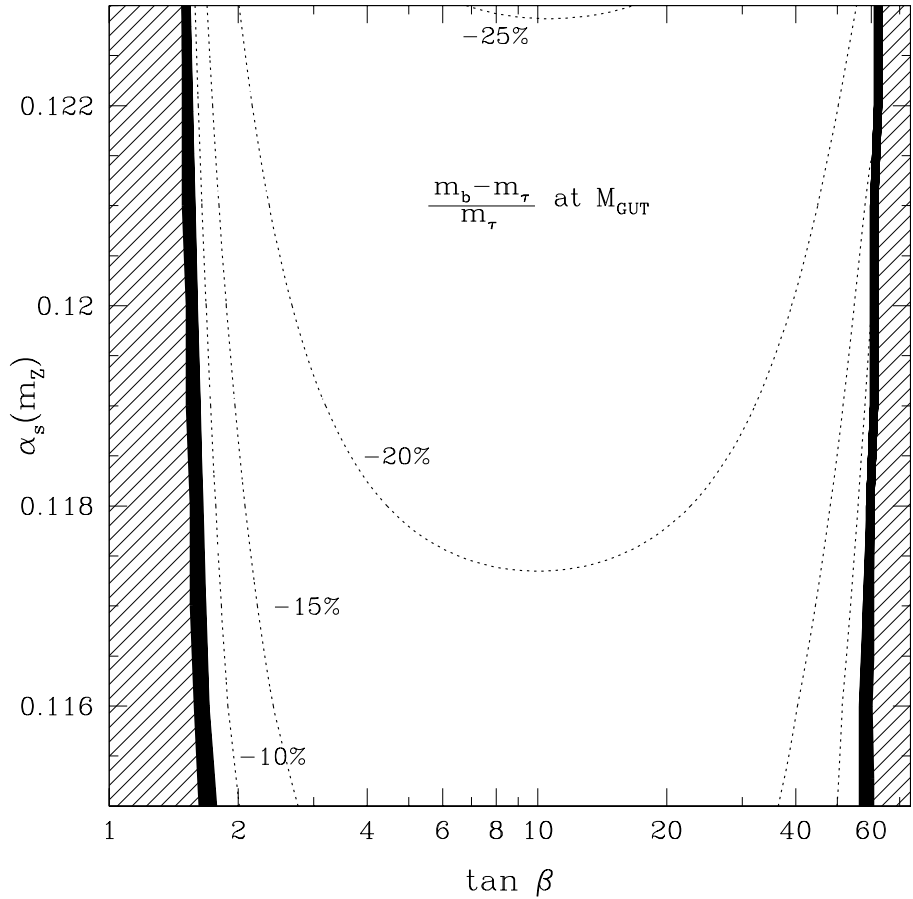}
\caption{Mismatch between the measured $b$-- and $\tau$--Yukawa
  couplings at the GUT scale. The
  hatched regions to the left and right are disallowed by
  perturbativity constraints. The dark regions exhibit approximate
  Yukawa unification. In the central regions, the contours label
  values of $(y_b-y_\tau)/y_\tau$, evaluated at the GUT scale.
}
\label{fig:gut}
\end{figure}

However, explicit calculations of Yukawa unification, within the context of the
MSSM and assuming a ``grand desert'' between the SUSY and GUT scales, find that
$b-\tau$ unification does not occur for generic values of
$\tan\beta$~\cite{als}. In
fact, for most $\tan\beta$, one finds that the physical $m_b$ predicted by
unification is much larger than measured; alternatively, given the measured
$m_b$, one find $y_b<y_\tau$ at the unification scale. Only at special values
of $\tan\beta$ does $b-\tau$ unification match experiment. These special values
correspond either to $y_t$ pseudo--fixed points 
($m_t^{\rm pole}/\sin\beta\simeq205\gev$)
or $y_b$ and $y_\tau$ pseudo--fixed points ($\tan\beta\simeq60$). For all other
values of $\tan\beta$, one generally finds:
\beq
0.75\lsim\left.\frac{y_b}{y_\tau}\right|_{GUT}\lsim0.85 \quad\quad
\Longleftrightarrow
\quad\quad -0.25\lsim\frac{m_{b,{\rm exp}}-m_{b,{\rm pred}}}{m_{b,{\rm
      pred}}}\lsim -0.15 ~.\nonumber
\eeq
We have shown this graphically in Fig.~\ref{fig:gut}
by plotting the difference between the
$b$-- and $\tau$--Yukawa couplings at the GUT scale as a function of
$\tan\beta$
and $\alpha_s(\mz)$. (For the plot, we have used $m_b(m_b)=4.2\gev$.) At very
large
and very small values of $\tan\beta$, it is impossible to talk about
unification because the theory is non--perturbative (the hatched
region). In the small regions near
the edge of perturbativity (the dark regions),
one does indeed find rough unification of $m_b$ and $m_\tau$ in the MSSM.
However, throughout the whole central region of the plot $b-\tau$ unification
fails, usually by 15--25\% as shown in the contours.
%%%%%%%%%%%%%%%%%%%%%%%%%%%%%%%%%%%%%%%%%%%%%%%%%%%%%%%%%%%%%%%%%%%
%%%%%%%%%%%%%%%%%%%%%%%%%%%%%%%%%%%%%%%%%%%%%%%%%%%%%%%%%%%%%%%%%%%

Thus, if $b-\tau$ unification is to survive, we must either live at the
pseudo--fixed point of some Yukawa coupling, or the one--loop
corrections to the
$b$--quark mass must {\sl reduce} $m_b$ to agree with the low measured value.
Thus minimal GUT unification implies $-25\%\lsim\delta m_b/m_b\lsim-15\%$
typically.

With this prediction in hand, we can go back to Figs.~2 and 3.
In Fig.~\ref{fig:hdecays}, we
see that in the region of interest of $\delta m_b/m_b$, large deviations in
$\rbtau$ can be expected. Even for $m_\hA=300\gev$, we can expect shifts
approaching +15\%. If $\rbtau$ is normalized experimentally by the $\hl\to\bar
bb$ branching ratio, then shifts in $\rbtau$ greater than zero correspond to
suppressed $\hl\to\tau^+\tau^-$ branching ratios.

In Fig.~\ref{fig:adecays},
we find that GUT--motivated shifts in the $b$--mass correspond to
shifts in $\rbtaua$ of nearly 80\%. Normalizing to the $A^0,H^0\to\bar bb$
branching ratios, we now find a suppression of the $A^0,H^0\to\tau^+ \tau^-$
branching ratios of nearly half. Thus GUTs would seem to prefer relatively
large shifts in $\rbtau$ and $\rbtaua$, with definite signs
corresponding to suppressed decays to $\tau$'s relative to $b$'s.

This represents then a rare opportunity to probe GUT physics more carefully,
and in the unexpected regime of Higgs decays. If the Higgs branching ratios are
found to shift, and to
do so with signs and magnitudes consistent with $b-\tau$ unification, this
would be additional circumstantial evidence in favor of a real unified gauge
group. On the other hand, large shifts in the wrong direction would certainly
constitute an argument against the simpler classes of unified models.

\section{Experimental comparison}

In order to experimentally detect deviations of $\rbtau$ from its SM
value, the SM value itself must be well--understood. To do so requires
careful measurement of three parameters: $m_b$, $m_\tau$ and
$\alpha_s$. Of these, $m_\tau$ is already extremely well--measured and requires
no further discussion. $\alpha_s$ has been measured over the last several years
with increasing precision. Global fits to data over a wide range of
energies gives $\alpha_s(\mz)=0.119\pm0.002$~\cite{pdg}.  Since $\alpha_s$
enters primarily as a 25\% radiative correction in $\rbtau$, the error due
to the uncertainty in $\alpha_s$ is small.

It is $m_b$ itself which is hardest to determine. The most direct,
though least precise, experimental determination of $m_b$ is through
three jet heavy quark production at LEP, where the effects of the
small $m_b$ are enhanced by the details of the jet clustering
algorithms. An
analysis by DELPHI gives $m_b(\mz)=2.67\pm0.50\gev$~\cite{delphi},
which translates to 
$m_b(m_b)=3.91\pm0.67$. The $\Upsilon$ system provides another
clean experimental measurement, though one in which theoretical
effects are harder to disentangle. Using QCD moment sum rules for
inclusive $b$--production in $e^+e^-$ collisions, and assuming that the
$\Upsilon$ saturates the higher moments, a recent value
of $m_b(m_b)=4.13\pm0.06\gev$~\cite{jamin} 
has been extracted. The latest
lattice extraction yields $m_b(m_b)=4.15\pm0.20\gev$~\cite{gimenez}.
The theoretical errors in these estimate seem to
be hard to quantify, a conservative range of $4.1\gev<m_b(m_b)<4.4\gev$
has been quoted in Ref.~\cite{pdg}, roughly an uncertainty of $\pm3.5\%$.

While the LEP--derived values of $\alpha_s$ are actually calculated at
the $Z$--pole (and then run down to $Q=m_b$ for comparison), the values
derived from heavy meson systems need to be run from $Q=m_b$ up to
$Q=\mz\simeq m_h$ for use in Eq.~(\ref{sm}). At one--loop, the
QCD renormalization group equations for $m_b$ can be solved:
\beq
m_b(\mz)=m_b(m_b)\left(\frac{\alpha_s(\mz)}{\alpha_s(m_b)}\right)^{12/23}\
{}.
\eeq
Thus, if $m_b(m_b)$ were known with infinite precision, there
would still be a 2\% uncertainty in $\rbtau$ from the current uncertainty
in $\alpha_s(\mz)$. However, that uncertainty is presently overwhelmed
by the uncertainties in $m_b$ itself. Thus an important aim of future
experimental and theoretical work should be to get the errors on $m_b$
down to the 2\% level. Given that the present error on $m_b(m_b)$ is
about 3-4\%, such an improvement does not appear to be beyond reach, especially
with forthcoming experimental efforts at the $B$--factories and theoretical
efforts
in lattice gauge theories. Only when the uncertainties in the SM
prediction for $\rbtau$ are below the few percent level will an
unequivocal measurement of $\delta m_b/m_b$ be possible in $\hl$
decays for a large portion of the parameter space.
Barring that, we must wait for the discovery of the $\hh$ and/or
$A^0$ where the effects can be expected to be much larger.  
Of course, if $\delta m_b/m_b$ is large while $m_A$ remains under a
few hundred GeV, observation of deviations in $h^0$ decays will be
possible even with the current precision in $m_b$.

\section{Conclusions}

We have shown that in the MSSM, unlike non--SUSY 2HDSMs, there can be
significant shifts in the ratio $BR(\phi\to\bar bb)/BR(\phi\to\tau^+\tau^-)$
for $\phi=\hl, \hh$ or $A^0$. These shifts in the $\hl$ decays may be
our first indication of SUSY, long before SUSY partners, or the additional
Higgs bosons, themselves are discovered. There is also a strong
correspondence between the shifts in the Higgs branching ratios and
Yukawa unification in minimal GUT models. In this way, significant
departures of the Higgs branching ratios away from their SM values
could teach us simultaneously about SUSY {\it and\/} GUT physics.

\section*{Acknowledgments}

The work of KSB is supported by funds from the Oklahoma State
University. CK is supported by the 
Department of Energy under contract DE--AC03--76SF00098.


\begin{thebibliography}{99}

\bibitem{hhg}
      For a review of the Higgs sector of the MSSM at tree--level, see
J.~Gunion, H.~Haber, G.~Kane and S.~Dawson, {\sl The Higgs Hunter's Guide},
Addison--Wesley, 1990.

\bibitem{hmass}
      H.~Haber and R.~Hempfling, \PRL{66}{91}{1815}; \\
      Y.~Okada, M.~Yamaguchi and T.~Yanagida, \PTP{85}{91}{1}; \\
      J.~Ellis, G.~Ridolfi and F.~Zwirner, \PLB{257}{91}{83};\\
      A.~Yamada, \PLB{263}{91}{233}.
\bibitem{maekawa}
      N.~Maekawa, \PLB{282}{92}{387}.

\bibitem{bkmw}
      K.S.~Babu, C.~Kolda, J.~March-Russell and F.~Wilczek, {\tt
hep-ph/9804355}, Phys.\ Rev.\ {\bf D} {\sl (in press)}.

\bibitem{gorishny}
      S.~Gorishny {\it et al.\/}, \MPLA{5}{90}{2703}.

\bibitem{edm}
      J.~Ellis, S.~Ferrara and D.~Nanopoulos, \PLB{114}{82}{231};\\
      W.~Buchm\"uller and D.~Wyler, \PLB{121}{83}{321};\\
      J.~Polchinski and M.~Wise, \PLB{125}{83}{393}.

\bibitem{banks}
      T.~Banks, \NPB{303}{88}{172};\\
      E.~Ma, \PRD{39}{89}{1922};\\
      R.~Hempfling, \PRD{49}{94}{6168}.

\bibitem{hrs}
      L.~Hall, R.~Rattazzi and U.~Sarid, \PRD{50}{94}{7048}.

\bibitem{lw}
      W.~Loinaz and J.~Wells, {\tt hep-ph/9808287}; \\
      M.~Carena, S.~Mrenna and C.~Wagner, {\tt hep-ph/9808312}.

\bibitem{ceg}
      M.~Chanowitz, J.~Ellis and M.~Gaillard, \NPB{128}{77}{506};\\
      A.~Buras, J.~Ellis, D.~Nanopoulos and M.~Gaillard, \NPB{135}{78}{66}.

\bibitem{als}
      See, for example: \\
      V.~Barger, M.~Berger and P.~Ohmann, \PRD{47}{93}{1093};\\
      M.~Carena, S.~Pokorski and C.~Wagner, \NPB{406}{93}{59};\\
      P.~Langacker and N.~Polonsky, \PRD{49}{94}{1454}.

\bibitem{pdg}
      C.~Caso {\it et al.\/} (Particle Data Group), \EPJC{3}{98}{1}.

\bibitem{delphi}
      P.~Abreu {\it et al.\/} (Delphi Collaboration), \PLB{418}{98}{430}.

\bibitem{jamin}
      M.~Jamin and A.~Pich, \NPB{507}{97}{334}.

\bibitem{gimenez}
       V.~Gim\'enez, G.~Martinelli and C.~Sachrajda, Nucl.\ Phys.\
       Proc.\ Suppl.\ {\bf 53}, 365 (1997).

\end{thebibliography}
\end{document}